\documentclass[showpacs,twocolumn,pre,floatfix]{revtex4}
\usepackage{psfrag,epsfig,amsfonts,amssymb,amsmath}
\usepackage{dcolumn}

\newcommand{\ttt}{{\cal T}}

\begin{document}

\title{Suppression of thermally activated escape by heating}

\author{Sebastian Getfert}
\author{Peter Reimann}
\affiliation{Universit\"at Bielefeld, Fakult\"at f\"ur Physik, 33615 Bielefeld, Germany}

\begin{abstract}
The problem of thermally activated escape over a potential 
barrier is solved by means of path integrals for one-dimensional 
reaction dynamics with very general time-dependences.
For a suitably chosen, but still quite simple static 
potential landscape, the net escape rate may be substantially 
reduced by temporally increasing the temperature above its 
unperturbed, constant level.
\end{abstract}
                                                 

\maketitle

Thermally activated escape over potential barriers is of 
relevance in a large variety of physical, chemical, 
and biological contexts \cite{han90}.
In the most common case, a potential barrier $\Delta U$ much 
larger than the thermal energy $kT$ yields an escape
rate exponentially small in $\Delta U/kT$.
A first major generalization, of importance for conceptual reasons
as well as due to numerous applications, are periodically modulated 
potentials \cite{gra84,mai96,dyk99,leh00,ber04},
resulting in a renormalization of $\Delta U$ which depends in
a very complicated manner on the details of the model.
In all these cases the rate is thus a very rapidly increasing
function of the temperature $T$.
In our present work, the main focus is on the complementary case 
of a static potential and a time dependent temperature \cite{ber04,rei96}.
In particular, we will demonstrate that in a suitably chosen, but
still fairly simple and generic potential landscape, the escape rate
of the unperturbed system at constant temperature may {\em decrease}
upon temporally
{\em increasing} the temperature.
In view of the above mentioned results for constant 
temperature, this is a quite unexpected and 
counter-intuitive result.
Indeed, given that thermal noise is indispensable to
escape, one would expect that an ``extra dose'' of noise
should always enhance escape.
Somewhat reminiscent previous findings 
always concern quite different types of systems:
Dissipative quantum tunneling in the deep cold \cite{gol92},
activationless electron transfer 
\cite{bix99},
complex protein dynamics near the solvent 
glass transition \cite{ans94},
models without a barrier against deterministic escape \cite{man96},
or non-dynamical systems \cite{vil01}.

As our main tool, we put forward a new path integral 
approach, which unifies and extends several related 
approximations 
\cite{gra84,mai96,dyk99,leh00,ber04,car81,shn97}.
Briefly, in different parameter regimes of the 
temporal modulations, the most relevant escape 
paths are of quite different character. 
Therefore, each regime was so far
treated separately and the crossover omitted.
Here, all potentially relevant paths are
represented in terms of a suitable, 
one-dimensional parametrization and 
are kept till the final rate formula
via an integral over all of them.

We consider the overdamped 1D Langevin equation
\begin{equation}
\eta(t) \dot x(t) = - U'(x(t),t) + \sqrt{2 \eta(t) kT(t)}\ \xi(t)
\label{1}
\end{equation}
with time-dependent friction $\eta(t) >0$, temperature $T(t)>0$,
and potential $U(x,t)$.
Dot and prime indicate temporal and spatial derivatives, 
$k$ is Boltzmann's constant,
and thermal fluctuations are modeled as usual \cite{han90} 
by $\delta$-correlated Gaussian noise $\xi(t)$.
For $T\to 0$, the deterministic dynamics
is required to exhibit exactly one stable 
orbit (attractor) $x_s(t)$ and one unstable
orbit (repeller) $x_u(t) > x_s(t)$.
Our main interest concerns the noise induced
transitions of $x(t)$ across $x_u(t)$
for small but finite temperatures $T(t)$,
quantified by the rate $\Gamma (t):=-\dot n(t)$
at which the probability $n(t)$ that 
$x(t)\leq x_u(t)$ changes in time.

To avoid unnecessary complications, we 
focus on initial conditions $x(t_0)=x_s(t_0)$,
and we require the existence of 
$D:=\lim_{t\to\infty}\int_0^t \frac{d\tau }{t} b(\tau )$ 
with $b(t):=kT(t)/\eta(t)$.
Next, we divide (\ref{1}) by $\eta(t)$
and employ transformed times 
$\tilde t(t):=\int_0^t d\tau \, b(\tau )/D$,
positions $\tilde x(\tilde t):=x(t(\tilde t))$,
and forces $\tilde F(x,\tilde t):=-D\, U'(x,t(\tilde t))/kT(t(\tilde t))$,
yielding, after dropping again the tildes,
\begin{equation}
\dot x(t) = F(x(t),t) + \sqrt{2D}\ \xi(t) \ .
\label{2}
\end{equation}
In the general formalism, we will work with (\ref{2}),
while 
specific examples will refer to (\ref{1}).
The corresponding (back-)transformation of the rates 
$\Gamma (t) = b(t) \tilde \Gamma (\tilde t(t))/D$
readily follows from the obvious transformation of the 
probabilities $\tilde n (\tilde t)= n(t(\tilde t))$.

We first recall some basics, previously derived and 
discussed in detail in Refs. \cite{leh00}:
For any given initial condition $x(t_0)=x_0$,
the probability density to find the stochastic process (\ref{2}) 
at any ``final'' time $t_f>t_0$ at the position $x_f$  
can be represented as path-integral
\begin{equation}
\rho (x_f, t_f\, | \, x_0, t_0 ) = 
\int_{x(t_0)=x_0}^{x(t_f)=x_f} {\cal D}x(t)\ e^{-S[x(t)]/D} \ ,
\label{3}
\end{equation}
with action $S[x(t)] := \int_{t_0}^{t_f} dt\ [\dot x(t) - F(x(t),t)]^2 /4$.
Once this formal integral is evaluated, the rate follows as
\begin{equation}
\Gamma (t) = -D\, \partial \rho (x_u(t),t \, |\, x_s(t_0),t_0)/\partial\, x_u(t) \ .
\label{4}
\end{equation}
For small $D$, the integral (\ref{3})
is dominated by the path $q(t)$ which minimizes the
action $S[x(t)]$ and thus satisfies the Euler-Lagrange equation
\begin{eqnarray}
\dot p(t) = -p(t)\, F'(q(t),t)   \ , \ \ 
p(t):=\dot q(t) - F(q(t),t)
\label{5}
\end{eqnarray}
with boundary conditions $q(t_0)=x_0$ and $q(t_f)=x_f$.
Accounting for all paths $x(t)$ ``close'' to $q(t)$
by means of a functional saddle point approximation 
in (\ref{3}) yields
\begin{equation}
\rho (x_f, t_f\, | \, x_0, t_0 )
=
[4\pi D Q(t_f)]^{-\frac{1}{2}}\, 
e^{- S[q(t)]/D}
\label{6}
\end{equation}
where $Q(t)$ satisfies $Q(t_0)=0$, $\dot Q(t_0)=1$, and
\begin{eqnarray}
\ddot Q(t) = 
\frac{d}{dt}[2\, Q (t)\, F'(q(t),t)] -
Q (t)\, p (t)\, F''(q(t),t)  \ .
\label{7}
\end{eqnarray}

Eqs. (\ref{4}), (\ref{6}) yield an approximation to
the rate $\Gamma(t)$, which in principle becomes
asymptotically exact as $D\to 0$ for any given 
$t>t_0$ with a unique absolute minimum of 
$S[x(t)]$, which generically is the case.
But under many circumstances of foremost interest
(e.g. relatively large $t-t_0$)
even fairly small $D$ are still far
from this asymptotic regime, i.e.
the saddle point approximation (\ref{6})
does not properly account for all relevant 
paths in (\ref{3}):
Basically, a typical escape 
path $x(t)$ spends almost all its time near 
$x_s(t)$, then crosses over into the vicinity 
of $x_u(t)$, and remains there for the rest 
of its time.
Any other behavior would 
yield a much larger action $S[x(t)]$ 
and thus is negligible in (\ref{3}). 
However, rather different cross over 
``time windows'' may still lead to almost 
equal $S[x(t)]$, and a simple saddle point
approximation is unable to properly account 
for such quite remote regions in path
space.
In some cases, there may exists further 
local minima of $S[x(t)]$ and additional
saddle point approximations around each of 
them may save the case \cite{mai96,leh00}.
The remaining problem is to keep track of 
all relevant minima and not 
to double count their neighborhoods 
if they get too close in path space.
In other cases, e.g. for $t$-independent 
$\eta$, $T$, and $U$ in (\ref{1}), 
there is a continuous  ``soft direction'' 
in path space, invalidating plain saddle 
point methods altogether \cite{car81}.

To overcome these problems we
impose on top of the boundary
conditions $x(t_0)=x_s(t_0)$ and
$x(t_f)=x_u(t_f)$ 
the extra condition that $x(t)$ arrives at some intermediate
point $x_i$ at a given time $t_i$, 
and in the end integrate over all $t_i\in [t_0, t_f]$
\cite{f1}.
The 
pertinent formal relation, satisfied by 
the conditional probability density (\ref{3}), is
\begin{equation}
\rho (x_f, t_f |  x_0, t_0 ) = 
\int\limits_{t_0}^{t_f} dt_i\, 
\rho (x_f, t_f |  x_i, t_i ) \,
\Psi_{x_i} \! (t_i |  x_0, t_0 )
\label{8}
\end{equation}
where $\Psi_{x} (t\, | \, x_0, t_0 )$ 
denotes the first passage time density 
across $x$, given $x(t_0)=x_0$.
For simplicity only, we assume from now on 
that $x_i$ is located 
well in between $x_s(t)$ and $x_u(t)$ 
and is $t$-independent.
Then, all non-negligible paths in (\ref{3}) 
starting from $x(t_i)=x_i$ must immediately 
cross over to $x_u(t)$ and thus admit 
for $\rho (x_f, t_f |  x_i, t_i )$ in (\ref{8})
a saddle point approximation (\ref{6}) 
free of all the above mentioned problems.
Focusing on $x_0=x_s(t_0)$ according to 
(\ref{4}), an analogous approximation (\ref{6})
holds for  $\rho (x_i, t_i |  x_0, t_0 )$,
since all relevant paths in (\ref{3}) now may 
leave the vicinity of $x_s(t)$ only in the very end.
By definition, $\rho (x_i, t_i |  x_0, t_0 ) d x_i$
is the probability that $x(t)$ from
(\ref{2}) is encountered within $[x_i,x_i+dx_i]$ at time
$t_i$, given $x(t_0)=x_0=x_s(t_0)$.
Most such $x(t)$ closely resemble the
most probable path $q(t)$ connecting
$q(t_0)=x_0$ with $q(t_i)=x_i$.
On the other hand $\Psi_{x_i} \! (t_i |  x_0, t_0 ) dt_i$
is the probability that $x(t)$ crosses $x_i$
for the first time during $t\in[t_i,t_i+dt_i]$.
It seems reasonable to guess that most such $x(t)$
once again closely resemble $q(t)$.
Hence, $\rho (x_i, t_i |  x_0, t_0 ) d x_i$ will 
essentially account for the same ``events'' as 
$\Psi_{x_i} \! (t_i |  x_0, t_0 ) dt_i$
provided we relate the considered intervals $dx_i$ 
and $dt_i$ via $dx_i=\dot q(t_i) dt_i$. 
Up to finite-$D$ corrections we thus obtain 
\begin{equation}
\Psi_{x_i} \! (t_i |  x_0, t_0 ) =  \dot q(t_i)\, \rho (x_i, t_i |  x_0, t_0 ) \ ,
\label{9}
\end{equation}
where $q(t)$ satisfies $q(t_0)=x_0=x_s(t_0)$, 
$q(t_i)=x_i$, and (\ref{5}).
More rigorously, our key relation (\ref{9}) follows by 
adapting Ref. \cite{dur92} to evaluate 
the derivative by $x_f$ of 
(\ref{8}) in the limit $x_f\to x_i$.
Details will be given elsewhere.

The evaluation of the escape rate (\ref{4}) by means of 
(\ref{8}), (\ref{9}), and (\ref{6}) is the first main result 
of our present work.
Similarly as in \cite{mai96,dyk99,leh00,ber04}, closed analytical
solutions of the concomitant differential equation (\ref{5}),
(\ref{7}) are only possible for special $F(x,t)$.
To this end, we focus on piecewise parabolic potentials
$U(x,t)$ in (\ref{1}), corresponding to piecewise linear
force fields in (\ref{2}) of the form
\begin{eqnarray}
F(x\leq 0 , t )  & = & \lambda_s(t) (x-y_s(t)) + f(t)
\nonumber
\\
F(x > 0 , t )  & = & \lambda_u(t) (x-y_u(t)) + f(t)
\label{10}
\end{eqnarray}
with $\lambda_s(t)\, y_s(t)=\lambda_u(t)\, y_u(t)$
(continuity at $x=0$).
Further, the existence of stable and unstable orbits
with $x_s(t)<x_i$ and $x_u(t)>x_i$ is required,
in particular 
$y_s(t),\,\lambda_s(t) <0 $, $y_u(t),\,\lambda_u(t) > 0$.
For the natural choice $x_i=0$,
a straightforward but somewhat tedious calculation
\cite{leh00,ber04} then yields for the rate (\ref{4}) the result
\begin{eqnarray}
\Gamma(t) & = & \int_{t_0}^t
d\tau  \ \frac{Z(t,\tau ,t_0)}{D}\ e^{-\Phi(t,\tau ,t_0)/D}
\label{11}
\\
\Phi(t,\tau ,t_0) & := & 
\frac{x_u^2(\tau )}{4\, I_u(\tau ,t)} + \frac{x_s^2(\tau )}{4\, I_s(\tau ,t_0)}
\label{12}
\\
Z(t,\tau ,t_0)  & := & 
\frac{[Y(\tau ,t_0) - x_s(\tau )]\, x_u(\tau )}{8 \pi [I_u(\tau ,t)\, I_s(\tau ,t_0)]^{3/2}}\, e^{\Lambda_u(\tau ,t)}
\label{13}
\\
Y(\tau ,t_0)  & := &  
I_s(\tau ,t_0)\, [f(\tau )- y_s(\tau )\, \lambda_s (\tau )]
\label{14}
\\
\Lambda_{s,u}(t,\tilde t)  & := & 
2\, \int_{\tilde t}^t d\tau \ \lambda_{s,u}(\tau)
\label{15}
\\
I_{s,u}(t,\tilde t)  & := & 
\left| \int_{\tilde t}^t d\tau\ e^{\Lambda_{s,u}(t,\tau)}\right|
\label{16}
\end{eqnarray}

We have verified that previous findings for time-periodic
\cite{mai96,leh00,ber04} and time-independent systems 
(\ref{1}) \cite{car81} are recovered as special cases.
Those from \cite{dyk99} are formally similar
but contain quantities (called ${\cal E}$ and $s(\phi)$) 
which are not explicitly available in general.

\begin{figure}
\epsfxsize=0.95\columnwidth
\epsfbox{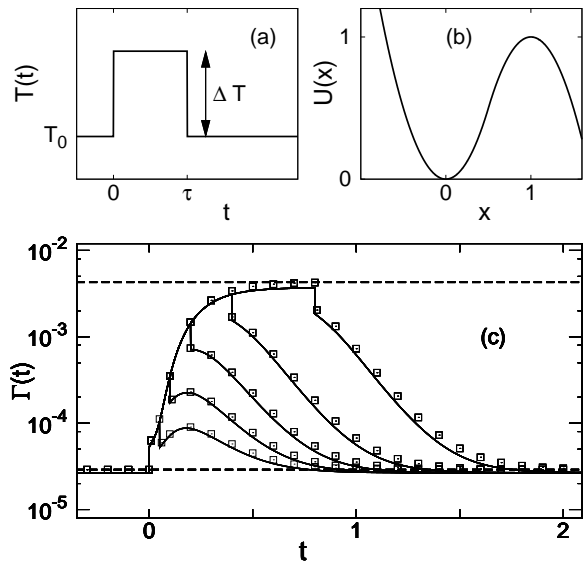} 
\caption{(a) Temperature pulse with parameters 
$T_0$, $\Delta T$, and $\tau$.
(b) Piecewise parabolic potential
with a well at $x_s=0$, a barrier at $x_u=1$,
curvatures $U''(x_s)=4$, $U''(x_u)=-4$,
and barrier height 
$\Delta U=U(x_u)-U(x_s)=1$.
(c) Time dependent escape rates for
$kT_0=k\Delta T=0.1$, $\eta(t)\equiv 1$, and 
$\tau=0.05$, $\tau=0.1$,
$\tau=0.2$, $\tau=0.4$, $\tau=0.8$
(bottom up).
Squares: Precise numerical solutions of
(\ref{1}) with 
seed $x(t_0=-5)=x_s$.
Solid lines:
Analytical approximation (\ref{11}) for the
equivalent transformed dynamics (\ref{2}), 
(\ref{10}).
Dashed lines:
Kramers rates \cite{han90} for  
$T=T_0$ (bottom) and $T=T_0+\Delta T$ (top).}
\label{fig1}
\end{figure}

As a first example we consider the dynamics (\ref{1}) 
with constant friction $\eta(t)\equiv 1$, a temperature
pulse $T(t)$ 
according to Fig. 1a,
and a static, piecewise parabolic potential $U(x)$,
see Fig. 1b.
Already for the moderately small 
temperatures from Fig. 1c,
the accuracy of the analytical approximation 
(\ref{11}) is quite good.
We found that it quickly improves even 
further upon decreasing temperatures.
After initial transients 
(omitted in Fig. 1c), 
Kramers rate is recovered until the temperature pulse 
sets in at $t=0$.
Then, the rate rapidly increases and
approaches the Kramers rate corresponding 
to $T_0+\Delta T$,
provided the pulse lasts sufficiently long.
Finally, an analogous relaxation back to
the original Kramers rate follows.
Discontinuities of $T(t)$ entail
jumps of $\Gamma (t)$.
While the initial transients are 
well understood \cite{shn97},
to the best of our knowledge no previously 
existing analytical approximation 
would be able to faithfully describe the
``perturbed and interfering transients'' for 
largely arbitrary pulses and pulse-sequences.

\begin{figure}
\epsfxsize=0.95\columnwidth
\epsfbox{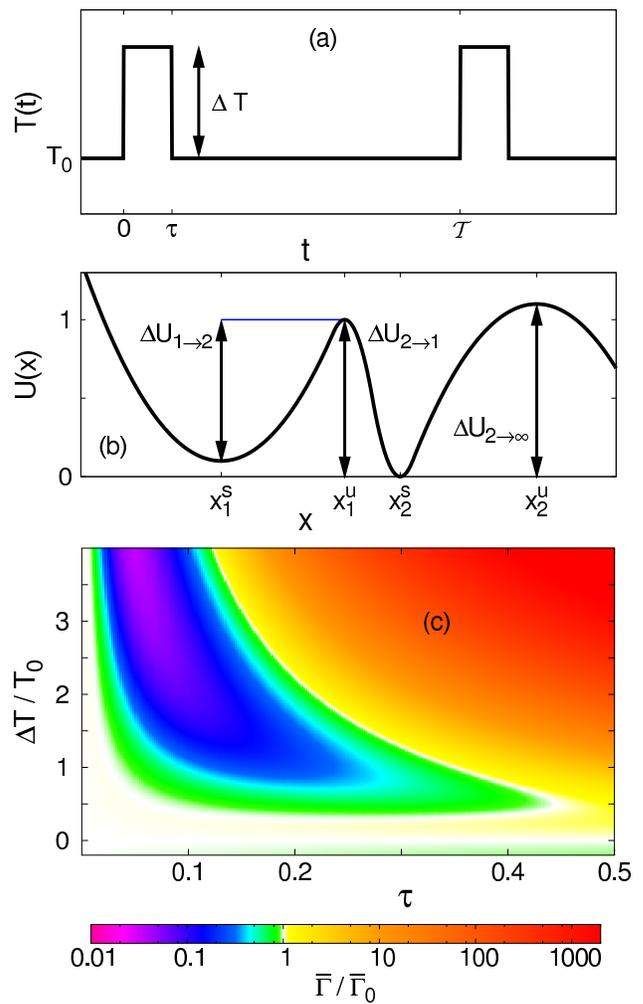} 
\caption{(a) Periodic temperature pulses with parameters 
$T_0$, $\Delta T$, $\tau$, and $\ttt$.
(b) Piecewise parabolic potential
with wells at 
$x^s_1=0$,
$x^s_2=2.04$, 
barriers at
$x^u_1=1.41$,
$x^u_2=3.6$, 
curvatures 
$U''(x^s_1)=1$, 
$U''(x^u_1)=-10$, 
$U''(x^s_2)=10$, 
$U''(x^u_2)=-1$, 
and barrier heights
$\Delta U_{1\to 2} = 0.9$,
$\Delta U_{2\to 1} = 1$,
$\Delta U_{2\to\infty} = 1.1$.
(c) Dependence of the 
effective rate $\bar\Gamma$ across $x_2^u$
on $\tau$ and $\Delta T/T_0$ in units of the 
rate $\bar\Gamma_0=9.11\cdot 10^{-11} $ for $T(t)\equiv T_0$.
Shown are analytical results for 
$\ttt=30$, $kT_0=0.05$, $\eta(t)\equiv 1$,
obtained as detailed in the main 
text.}
\label{fig2}
\end{figure}

Next, we consider (\ref{1}) with
a periodically pulsating temperature $T(t)$ and 
a piecewise parabolic potential $U(x)$ exhibiting
two barriers and two wells, see Fig. 2.
Transitions from 
$x_1^s$ to $x_2^s$ are described by 
the rate $\Gamma_{1\to 2}(t)$,
those from $x_2^s$ to $x_1^s$ by
$\Gamma_{2\to 1}(t)$,
and those from $x_2^s$ 
towards $x=\infty$ by 
$\Gamma_{2\to \infty}(t)$.
After suitable time- and space-transformations
(cf. (\ref{2})), each rate 
can be approximated 
according to (\ref{11}).
Since they are small (transitions are rare),
it is sufficient --
as far as the populations $n_1(t)$ and
$n_2(t)$ of the two wells are concerned --
to consider their averages over one 
period $\ttt$, denoted by
$\bar \Gamma_{1\to 2}$,
$\bar \Gamma_{2\to 1}$, and
$\bar \Gamma_{2\to \infty}$.
Then, the populations $\vec n(t):=(n_1(t),n_2(t))$
are governed by the master equation \cite{han90}
$\dot{\vec n}(t)=-M\vec n(t)$ with matrix elements
$M_{11}=-M_{21}=\bar \Gamma_{1\to 2}$,
$M_{12}=-\bar \Gamma_{2\to 1}$, and
$M_{22}=\bar \Gamma_{2\to 1}+\bar \Gamma_{2\to \infty}$.
The smallest eigenvalue of $M$ is denoted by 
$\bar\Gamma$ and represents the ultimate rate
of escape towards $\infty$ after initial 
relaxation processes, governed by the other
eigenvalue of $M$, have died out.
Since the two eigenvalues differ
by a huge, Boltzmann-Arrhenius-type 
factor, the total probability
$n_1(t)+n_2(t)$ that $x(t)<x_2^u$
is expected and numerically observed
to actually exhibit a practically 
perfect exponential decay
$e^{-\bar\Gamma t}$ for all $t>0$.
The analytical results for $\bar\Gamma$ 
are depicted in Fig. 2c.
Their agreement with our numerical
findings for the decay rate 
(not shown) is comparable to Fig. 1c.

The most striking feature of Fig. 2c is a
substantial {\em reduction} of the net 
escape rate $\bar\Gamma$
upon superimposing temperature pulses of suitable
duration $\tau$ and amplitude $\Delta T$
to the ``unperturbed'' temperature $T_0$.
Roughly speaking,
$\Gamma (t)$ in Fig. 1b
approaches the instantaneous 
Kramers rate the quicker, the
larger the curvatures in Fig. 1b
are \cite{shn97}.
Since the curvatures relevant for
$\bar\Gamma_{2\to 1}$ are larger than
those for $\bar\Gamma_{1\to 2}$ and 
$\bar\Gamma_{2\to \infty}$ (see Fig. 2b),
sufficiently small $\tau$ mainly
affect $\bar \Gamma_{2\to1}$
and thus lead to a reduction 
of the net decay rate $\bar\Gamma$.
We verified that already a single temperature 
pulse (Fig. 1a) indeed yields an analogous
reduction of escapes events.
Fig. 2c further shows that
the effect is overruled by
competing secondary effects
when $\tau$ and/or $\Delta T$
become too small.
Finally, we have obtained very similar
results also for $U(x^s_1)<U(x^s_2)$,
but from the viewpoint of equilibrium rates 
\cite{han90}, the case $U(x^s_1)>U(x^s_2)$ 
shown in Fig. 2 seems even more 
surprising to us.

Experimentally, potentials like in Fig. 2b 
are ubiquitous in the context of chemical 
reactions.
E.g. in the modified case $U(x^s_1)<U(x^s_2)$
these are reactions proceeding in two-steps 
via an intermediate (metastable state $x^s_2$).
Temperature pulses could be generated, 
among others \cite{rei96},
by means of a flashing black body 
radiator.
More realistic are short laser pulses \cite{her06},
whose basic effects (on the reacting molecules 
{\em and} their environment) may still be roughly 
modeled by a temperature pulse.
On the other hand, we expect that our above mentioned 
main finding will be qualitatively robust against 
various modifications of the pulsed perturbation, 
including more realistic models for tailored
laser pulses.
A further experimental playground are colloidal
particles in a suitably designed potential 
landscape by exploiting 
light \cite{bab04},
dielectrophoretic \cite{rou94},
or magnetic \cite{pam06}
forces.
Temperature pulses in the form of acousto-mechanical 
white noise may be generated by means 
of piezo elements \cite{cou07}.

Since temperature pulses in (\ref{1}) are basically
equivalent to potential modulations in (\ref{2}), 
we arrive at yet another quite astonishing
conclusion: The escape rate for a $t$-independent 
temperature and a potential landscape as in Fig. 2b
may {\em decrease} if the amplitude (multiplicative factor)
of the potential is temporally {\em reduced} without
any other change of its ``shape''.
This effect should be readily observable
with colloidal systems 
\cite{bab04,rou94,pam06} and possibly 
also with cold atoms in laser induced 
optical lattices \cite{gom08} or
in complex reaction networks \cite{mil96}.

\begin{center}
\vspace{-5mm}
---------------------------
\vspace{-4mm}
\end{center}
We thank B. Gentz for stimulating discussions.
This work was supported by DFG
under SFB 613


\begin{thebibliography}{99}
  
\bibitem{han90}P. H\"anggi, P. Talkner, and M. Borkovec, Rev. Mod. Phys. {\bf 62}, 251 (1990).

\bibitem{gra84} 
R. Graham and T. T\'el, J. Stat. Phys.  
{\bf 35}, 729 (1984); 
P. Jung, Phys. Rep. {\bf 234}, 175 (1993);
P. Talkner, New J. Phys. {\bf 1}, 4 (1999).

\bibitem{mai96}
R. S. Maier and D. L. Stein, Phys. Rev. Lett. {\bf 77}, 4860 (1996);
{\bf 86}, 3942 (2001)

\bibitem{dyk99}
V. N. Smelyanskiy, M. I. Dykman, and B. Golding, Phys. Rev. Lett. {\bf 82}, 3193 (1999);
D. Ryvkine and M. I. Dykman, Phys. Rev. E {\bf 72}, 011110 (2005);

\bibitem{leh00}
J. Lehmann, P. Reimann, and P. H\"anggi, Phys. Rev. Lett. {\bf 84}, 1639 (2000);
Phys. Rev. E {\bf 62}, 6282 (2000)

\bibitem{ber04}
N. Berglund and B. Gentz, J. Stat. Phys. {\bf 114}, 1577 (2004); 
Europhys. Lett. {\bf 70}, 1 (2005)

\bibitem{rei96}
P. Reimann, R. Bartussek, R. H\"aussler, and P. H\"anggi,
Phys. Lett. A {\bf 215}, 26 (1996);
J.~Luczka, T.~Czernik, and P.~H\"anggi, 
Phys. Rev. E {\bf 56}, 3968 (1997);
Y.-X. Li, 
Physica A {\bf 238}, 245 (1997);
Y.-D. Bao, 
Physica A {\bf 273}, 286 (1999);
Comm. Theor. Phys. {\bf 34}, 441 (2000);
P. Reimann, Phys. Rep. {\bf  361}, 57 (2002);
R. Eichhorn and P. Reimann,
Phys. Rev. E {\bf 70}, 035106(R) (2004);
N. Li, P. H\"anggi, and B. Li,
EPL {\bf 84}, 40009 (2008);
J. Iwaniszewski and A. Wozinski,
EPL {\bf 82}, 50004 (2008)

\bibitem{gol92}
B. Golding, N. M. Zimmerman, and S. N. Coppersmith,
Phys. Rev. Lett. {\bf68}, 998 (1992);
K. Chun and N. O. Birge, Phys. Rev. B {\bf48}, 11500 (1993).

\bibitem{bix99}
M. Bixon and J. Jortner,
Adv. Chem. Phys. {\bf 106}, 35 (1999).

\bibitem{ans94}
A. Ansari, C. M. Jones, E. R. Henry, J. Hofrichter, and W. A. Eaton,
Biochemistry {\bf 33}, 5128 (1994).

\bibitem{man96}
R. N. Mantegna and B. Spagnolo, 
Phys. Rev. Lett. {\bf 76}, 563 (1996);
P. Reimann, J. Stat. Phys. {\bf 82}, 1467 (1996)

\bibitem{vil01}
J. M. G. Vilar and J. M. Rubi, Phys. Rev. Lett. {\bf 86}, 950 (2001).

\bibitem{car81}
B. Caroli, C. Caroli, and B. Roulet,
J. Stat. Phys. {\bf 26}, 83 (1981).

\bibitem{shn97}
V. A. Shneidman, Phys. Rev. E {\bf 56}, 5257 (1997);
M. Bier, I. Der\'enyi, M. Kostur, and R. D. Astumian,
{\em ibid} {\bf 59}, 6422 (1999);
S. M. Soskin, V. I. Sheka, T. L. Linnik, and R. Mannella,
Phys. Rev. Lett. {\bf 86}, 1665 (2001)

\bibitem{f1}
Integrating over $x_i$ (keeping $t_i$ fixed)
would also be possible but not really solve 
the above mentioned problems.

\bibitem{dur92}
J. Durbin and D. Williams, J. Appl. Prob. {\bf 29}, 291 (1992)

\bibitem{her06}
I. V. Hertel and W. Radloff, Rep. Prog. Phys. {\bf 69}, 1897 (2006);
P. Nuernberger et al., Phys. Chem. Chem. Phys. {\bf 9}, 2470 (2007)

\bibitem{bab04}
D. Babic et al., 
Europhys. Lett. {\bf 67}, 158 (2004);
S. H. Lee et al., 
Phys. Rev. Lett. {\bf 94}, 110601 (2005);
S. Bleil et al., 
Phys. Rev. E {\bf 75}, 031117 (2007)

\bibitem{rou94}
J. Rousselet. et al., Nature {\bf 370}, 446 (1994);
L.~P. Faucheux and A.~Libchaber, 
J. Chem. Soc.  Faraday Trans. {\bf 91}, 3163 (1995);
L. Gorre-Talini et al., 
Chaos {\bf 8}, 650 (1998)

\bibitem{pam06}
N. Pamme,
Lab on a Chip {\bf 6}, 24 (2006)

\bibitem{cou07}
C. Coupier, M. Saint Jean, and C. Guthmann, 
EPL {\bf 77}, 60001 (2007)

\bibitem{gom08}
R. Gommers, V. Lebedev, M. Brown, and F. Renzoni, 
Phys. Rev. Lett. {\bf 100}, 040603 (2008)

\bibitem{mil96}
M. M. Millonas and D. R. Chialvo, Phys. Rev. Lett. {\bf 76}, 550 (1996);
C. R. Hickenboth et al. Nature {\bf 446}, 423 (2007);
A. Kargol and K. Kabza, Phys. Biol. {\bf 5}, 02003 (2008);
N. A. Sinitsyn and I. Nemenman, Phys. Rev. Lett. {\bf 99}, 220408 (2007);
S. Rahav, J. Horowitz, and C. Jarzynski, {\em ibid.} {\bf 101}, 140602 (2008)


\end{thebibliography}
\end{document}